\documentclass[fleqn,twoside,twocolumn,nofootinbib,showkeys]{revtex4-1} 

\usepackage[utf8]{inputenc}
\usepackage{amsmath}
\usepackage{amssymb}
\usepackage{dcolumn}
\usepackage{url}
\usepackage{hyperref}
\usepackage{verbatim}
\usepackage{listings}
\usepackage{fancyvrb}
\usepackage[breakable]{tcolorbox}
\usepackage{spverbatim}
\usepackage{xcolor}
\usepackage{lipsum}
\usepackage{float}
\usepackage{graphicx} 
\usepackage{booktabs}
\usepackage{float}

\begin{document}
\title{Mamba Quantum Eigensolver (MQE): A Selective State Space Model for Scalable Ground State Simulations in Quantum Chemistry}
\author{G. X. A. Petronilo}
\affiliation{
Centro Brasileiro de Pesquisas Físicas, Rua Dr. Xavier Sigaud 150,
22290-180 Rio de Janeiro, Rio de Janeiro, Brazil}
\email{gustavopetronilo@gmail.com}

\author{M. R. Araújo}
\affiliation{Quantum Industrial Innovation Universidade SENAI CIMATEC Salvador, BA, Brazil}

\author{Erico S. Teixeira}
\affiliation{Centro de Excelência em Computação Quântica, Venturus Centro de Inovação Tecnológica, Estrada Giuseppina Vianelli di Napolli, 1185, 13086-530, Campinas, SP, Brazil}

\author{Clebson Cruz}
\email{clebson.cruz@ufob.edu.br}
\affiliation{Centro das Ciências Exatas e das Tecnologias, Universidade Federal do Oeste da Bahia, Rua Bertioga 892, Morada Nobre, 47810-059}

\date{\today}

\begin{abstract}
Predicting molecular electronic structure using Quantum Computing approaches is a challenge in Computational Chemistry. Traditional Variational Quantum Eigensolvers (VQEs) are typically hindered by optimization challenges. The alternative Generative Quantum Eigensolver (GQE) mitigates this by reformulating ground-state search as discrete sequence generation, but its Transformer backbone suffers from quadratic computational complexity, bottlenecking scalability for larger molecules. To address this, we introduce the Mamba Quantum Eigensolver (Mamba-QE), which replaces the Transformer with a Selective State Space Model (SSM) to process operator sequences with linear $\mathcal{O}(L)$ complexity. Additionally, we integrate a distance-conditioned mechanism enabling the direct prediction of complete potential energy surfaces (PES). Evaluated on the $H_2$ molecule, Mamba-QE reproduces Full Configuration Interaction (FCI) reference energies well within chemical accuracy, achieving an energy error below $3\times10^{-6}$ Ha across 16 geometries with zero variational violations. Mamba-QE maintains high expressive power while establishing a robust, scalable foundation for quantum simulations of complex molecular systems. 
\end{abstract}

\maketitle




\section{Introduction}

Predicting the electronic structure of molecules is a fundamental challenge in computational chemistry, with direct implications for catalyst design\cite{10.1039/d5sc08985a}, materials discovery\cite{DL_R}, and pharmaceutical development\cite{slct.202507418}. Solving the electronic Schrödinger equation exactly is resource-demanding for most systems. Consequently, approximate methods have been developed, balancing computational cost against chemical accuracy. Nevertheless, many chemically relevant systems, particularly those with strong electron correlation, remain beyond scalable classical approximations, motivating the development of quantum algorithms that exploit quantum hardware to accelerate molecular simulations \cite{cao2019quantumchemistry}.

Among the most promising approaches for near-term quantum devices are Variational Quantum Algorithms (VQAs)\cite{Cerezo2021}, particularly the Variational Quantum Eigensolver (VQE) \cite{peruzzo2014variational}. This approach combines parameterized quantum circuits with a classical optimization loop to estimate molecular ground-state energies through the variational principle. The framework has demonstrated success for small molecules and has become one of the most extensively studied quantum algorithms for electronic structure calculations. However, its scalability is limited by well-known challenges, including barren plateaus\cite{Larocca2025}, optimization instability \cite{PhysRevA.107.032407, 9259985}, strong dependence on ansatz design\cite{pr14132095, 812w-ytnb}, and the increasing computational cost of the hybrid optimization loop as circuit depth and molecular complexity grow. These limitations motivate the search for alternatives that reduce or eliminate continuous parameter optimization.

A recent approach in this direction is the Generative Quantum Eigensolver (GQE) \cite{nakaji2024gqe}, which reformulates ground-state search as a generative sequence modeling problem. Instead of optimizing continuous parameters within a fixed quantum circuit, GQE trains a generative model to construct quantum circuits by sequentially selecting operators from a predefined gate pool. The generated circuits are evaluated on a quantum processor, and their measured energies are used as reward signals for reinforcement learning. By transferring the optimization process from a continuous parameter landscape to a discrete sequence generation task, GQE searches to reduce the barren plateau while enabling adaptive circuit discovery.

The original GQE employs a Transformer architecture inspired by GPT models \cite{vaswani2017attention,radford2019language} as its generative backbone. However, the self-attention mechanism in this architecture exhibits quadratic computational and memory complexity with respect to sequence length. As molecular systems become larger, longer operator sequences are required to represent quantum circuits, causing the computational cost of the Transformer backbone to grow rapidly. Consequently, the scalability of the classical sequence model itself becomes a bottleneck for extending GQE to larger molecular systems.

This motivates exploring alternative architectures that preserve the expressive power of Transformers while offering lower computational complexity. Recently, Selective State Space Models (SSMs), particularly the Mamba architecture \cite{gu2023mamba}, have emerged as a promising alternative. Rather than relying on self-attention, Mamba employs an input-dependent state-space formulation that processes sequences with linear computational complexity while maintaining competitive modeling capability across a broad range of sequence learning tasks. 

Building directly upon the GQE framework \cite{nakaji2024gqe}, we propose the \textbf{Mamba Quantum Eigensolver (Mamba-QE)}. Architecturally, Mamba-QE is a targeted enhancement of the GQE paradigm: it replaces the $\mathcal{O}(L^2)$ multi-head self-attention backbone of GPT-QE with a Mamba-based SSM, effectively scaling the sequence processing engine to linear $\mathcal{O}(L)$ complexity while preserving the core discrete-optimization reinforcement learning loop.


Leveraging this efficient linear-complexity backbone, we further extend GQE by introducing a continuous distance-conditioning mechanism, $\phi(R)$. While previous GQE implementations require training independent models for isolated molecular geometries, our geometric embedding concatenates interatomic spatial coordinates directly into the token representations. This allows a single unified Mamba-QE policy to generate tailored quantum circuits across an entire Potential Energy Surface (PES), transforming GQE from a static point-wise solver into a geometrically continuous generative framework.


From a chemical perspective, traditional physics-inspired ansätze, such as Unitary Coupled Cluster with Singles and Doubles (UCCSD)\cite{Romero_2019}, offer rigor but scale steeply in circuit depth ($O(N^4)$ or higher in terms of spin-orbitals). While generative frameworks mitigate this by sampling compact, problem-tailored operator sequences, Transformer-based backbones quickly hit an $\mathcal{O}(L^2)$ computational bottleneck as the required sequence length $L$ grows for larger active spaces and strongly correlated regimes.

By replacing the Transformer with a SSMs, Mamba-QE eliminates this bottleneck, maintaining linear $O(L)$ scaling. This efficiency is essential for scaling generative circuit discovery to complex chemical phenomena, such as transition-metal complexes, bond-breaking processes, and open-shell radicals, where capturing both static and dynamic correlation demands longer operator sequences and standard VQE optimizations collapse into barren plateaus.


Furthermore, analyzing the operator sequences generated by Mamba-QE provides direct physical interpretability. By tracking which quantum gates the model prioritizes across different geometries, one can map the onset of dynamic versus static correlation along a PES. Thus, Mamba-QE serves not only as a scalable solver for ground-state energies but also as a generative tool for deciphering the essential correlation structures of complex chemical systems.

As a proof-of-concept validation of this methodology, the proposed architecture is evaluated on the hydrogen molecule ($\text{H}_2$), where it reproduces Full Configuration Interaction (FCI) reference energies, achieving an energy error below $3\times10^{-6}$ Ha across 16 internuclear distances with zero variational violations. While benchmarks on $\text{H}_2$ do not yet expose hardware scaling limits, they establish the foundational empirical evidence that transitioning to a linear-complexity SSM retains high physical expressivity without degrading optimization stability, providing a scalable backbone for deeper quantum circuits in complex chemical systems.


Taken together, these developments establish Mamba-QE as a generative framework that connects efficient sequence modeling with geometry-aware quantum circuit generation for molecular electronic-structure calculations. The $\text{H}_2$ results provide a proof of concept for this approach and establish a basis for investigating its scalability to larger molecular systems.

The remainder of this paper is organized as follows. Section II presents the methodology, including the Generative Quantum Eigensolver framework, the Mamba-based architecture, the distance-conditioning mechanism, and the dynamic quality-threshold scheduling. Section III presents and discusses the numerical results for $\text{H}_2$. Finally, Section IV summarizes the main findings and discusses directions for future work.

\section{Methods}

\subsection{Generative Quantum Eigensolver and the GPT-QE Baseline}

The Generative Quantum Eigensolver (GQE) \cite{nakaji2024gqe} is a hybrid quantum-classical framework that recasts ground-state search as a discrete optimization problem solved via generative modeling. Rather than optimizing the continuous parameters of a fixed ansatz circuit, as in VQE \cite{peruzzo2014variational}, GQE trains a classical generative model to sample sequences of operators from a predefined gate pool $\mathcal{P} = \{U_1, U_2, \ldots, U_K\}$. The generative model therefore produces an ordered sequence of operator indices, $\vec{j}=(j_1,\ldots,j_L)$, which is then mapped to a quantum circuit by composing the corresponding operators,
\begin{equation}
    U_L(\vec{j}) = U_{j_L} \cdots U_{j_1}
\end{equation}
is applied to a reference state $|\rho_0\rangle$ to yield trial states whose energy expectation values serve as a reward signal during training. This formulation shifts the search from the highly non-convex continuous optimization of circuit parameters to the discrete space of circuit architectures.


The original implementation proposed by Nakaji et al. employs a decoder-only Transformer architecture, specifically a GPT-2-style model \cite{radford2019language} as the generative backbone, giving rise to the GPT-QE baseline. In GPT-QE, the generative model is trained autoregressively: at each step $t$, it computes a conditional probability distribution $p_\theta(j_t \mid j_1, \ldots, j_{t-1})$ over the operator pool and samples the next gate. The sequence probability factorizes as
\begin{equation}
    p_\theta(\vec{j}) = \prod_{t=1}^{L} p_\theta\!\left(j_t \mid j_1, \ldots, j_{t-1}\right),
\end{equation}
where $L$ denotes the sequence length. The model parameters $\theta$ are updated via a reinforcement learning objective using the expectation value $\langle \Psi(\vec{j}) | \hat{H} | \Psi(\vec{j}) \rangle$ as the cost function to minimize. The self-attention mechanism of the Transformer, operating with $\mathcal{O}(L^2)$ computational and memory complexity in the sequence length $L$, captures long-range dependencies between previously selected gates to inform future selections.

Despite the success of this approach in demonstrating the feasibility of generative circuit search for molecular electronic structure Hamiltonians \cite{nakaji2024gqe}, the quadratic scaling of the Transformer attention mechanism presents a bottleneck as the operator sequence length, and thus the circuit depth, grows with system size. For molecules requiring large active spaces, such sequences can become prohibitively long, making the computational cost of GPT-QE training a limiting factor for practical applications.

\subsection{Mamba-QE: A Selective State Space Model Alternative}

To address the scalability limitations of the Transformer backbone, we propose replacing the GPT architecture in GQE with the Mamba architecture \cite{gu2023mamba}, yielding the \textbf{Mamba-QE} framework. Mamba is a selective state space model (SSM) that processes sequences with linear complexity $\mathcal{O}(L)$ in the sequence length while retaining competitive modeling capacity relative to Transformers \cite{gu2023mamba}. The key innovation of Mamba over earlier SSMs such as S4 \cite{gu2022s4} is the introduction of input-dependent, content-aware gating: the parameters $(\mathbf{B}, \mathbf{C}, \Delta)$ are parameterized as functions of the current input, enabling the model to selectively propagate or discard information along the sequence at a substantially lower computational cost than multi-head self-attention.

Formally, for a hidden state dimension $d$ and an input sequence of operators $\vec{j} = (j_1, \ldots, j_L)$, let $x_t \in \mathbb{R}^{d}$ denote the continuous embedding vector of token $j_t$ at step $t$. The Mamba block evolves a latent state $\mathbf{h}_t \in \mathbb{R}^{d}$ according to the recurrence
\begin{equation}
    \mathbf{h}_t = \overline{\mathbf{A}}_t\,\mathbf{h}_{t-1} +
    \overline{\mathbf{B}}_t\, x_t, \qquad y_t = \mathbf{C}_t\,\mathbf{h}_t,
    \label{eq:mamba_recurrence}
\end{equation}
where $\overline{\mathbf{A}}_t$ and $\overline{\mathbf{B}}_t$ are discretized versions of the continuous-time SSM matrices. Crucially, $\mathbf{B}_t$, $\mathbf{C}_t$, and the timescale parameter $\Delta_t$ are input-dependent projections of $x_t$, whereas $\mathbf{A}$ remains static. This selectivity allows Mamba to model long-range dependencies in operator sequences without the quadratic memory overhead of self-attention, enabling hardware-aware parallel training via a associative scan algorithm while retaining an $\mathcal{O}(1)$ state memory overhead per step during autoregressive generation \cite{gu2023mamba}.

Mamba-QE employs a two-stage supervised generative refinement procedure consisting of an offline warm-up stage followed by iterative self-training with a quantum-evaluated replay buffer. Unlike policy-gradient implementations of GQE, the present implementation does not directly optimize the model through a reinforcement-learning gradient. Instead, quantum evaluations are used to construct additional training examples, which are subsequently incorporated into the supervised learning process.

During the warm-up stage, random operator sequences are generated independently for each molecular geometry. For each sequence, the energy of every intermediate subsequence is evaluated using the quantum simulator. If the generated operator sequence is \(\mathbf{j}=(j_1,\ldots,j_L)\), the resulting training target is the sequence of cumulative energies

\begin{equation}
\mathbf{E} =
(E_1,E_2,\ldots,E_L)
\end{equation}

where \(E_l\) corresponds to the expectation value obtained after applying the first \(l\) operators. The model receives the corresponding operator tokens together with the molecular geometry and learns to reproduce these subsequence energies through the cumulative token-logit representation.

For a sequence of tokens, the model produces a scalar logit \(w_t\) for the selected operator at each autoregressive position. The predicted energy after \(l\) operators is defined as

\begin{equation}
\hat{E}_l = \sum_{t=1}^{l} w_t
\end{equation}

The training objective is therefore the mean squared error between the cumulative predicted energies and the quantum-evaluated subsequence energies,

\begin{equation}
\mathcal{L}_{\mathrm{MSE}}=
\frac{1}{NL}
\sum_{i=1}^{N}
\sum_{l=1}^{L}
\left(\hat{E}^{(i)}_l-E^{(i)}_l\right)^2
\end{equation}

where \(N\) is the number of training sequences and \(L\) is the circuit sequence length. This objective corresponds directly to the implementation in which the selected-token logits are cumulatively summed before comparison with the subsequence-energy targets.

After the warm-up stage, Mamba-QE enters an iterative refinement phase. At predefined refinement iterations, the current model generates candidate operator sequences conditioned on randomly selected molecular geometries. Each generated sequence is evaluated with the quantum simulator, and its final energy is used to determine whether the sequence is admitted to the replay buffer. The buffer therefore converts quantum evaluations into additional supervised training data.

At each subsequent training iteration, the model is trained using a mixture of the original randomly generated dataset and samples drawn from the replay buffer. In the present implementation, the default buffer contribution during refinement is 30\% of each training batch, while the warm-up phase uses no replay-buffer samples. This procedure progressively shifts part of the training distribution toward lower-energy circuits while retaining randomly generated examples to preserve exploration.

The refinement process is consequently best interpreted as a self-training loop rather than conventional reinforcement learning: quantum evaluation determines which generated sequences become training examples, but no policy-gradient estimator or explicit reward-gradient update is employed.

\subsection{Selective state-space architecture}
The sequence model is implemented as a minimal selective state-space architecture. The model uses a hidden dimension \(d_{\mathrm{model}}=128\), six residual Mamba blocks, state dimension \(d_{\mathrm{state}}=16\), expansion factor of 2, and a depthwise convolution with kernel size 4. The resulting expanded state dimension is

\begin{equation}
   d_{\mathrm{inner}}=2d_{\mathrm{model}}=256 
\end{equation}

For each input representation \(x_t\), the Mamba block first applies an input projection that separates the state-space input from a residual gating branch. The state-space parameters are then obtained through an input-dependent projection,

\begin{equation}
    (\tilde{\Delta}_t,B_t,C_t) = W_x x_t+b_x
\end{equation}

where the first component is mapped to the model-dependent timescale through

\begin{equation}
    \Delta_t = \operatorname{softplus}
\left(W_\Delta\tilde{\Delta}_t+b_\Delta\right)
\end{equation}

The state matrix is parameterized as

\begin{equation}
   A=-\exp(A_{\log}) 
\end{equation}

where \(A_{\log}\) is learned. The selective state update is implemented through equation \ref{eq:mamba_recurrence}.

Thus, \(B_t\), \(C_t\), and \(\Delta_t\) depend on the current input representation, whereas the underlying \(A\) parameter is learned independently of the input. This input dependence provides the selective mechanism through which the model can dynamically regulate which information is propagated through the operator sequence.

For computational stability on longer sequences, the selective scan is processed in chunks of 64 positions. During training, gradient checkpointing is used rather than detaching the hidden state, thereby limiting activation memory while preserving gradient propagation across chunks.

\subsection{Distance-Conditioned Mamba Architecture}


To generalize the generative framework across varying molecular geometries, we introduce a geometric conditioning mechanism into the Mamba-QE architecture. This allows the model to learn a unified policy capable of generating tailored quantum circuits across an entire Potential Energy Surface (PES), rather than optimizing for a single static molecular configuration.

Concretely, for a given interatomic distance $R \in \mathbb{R}$, we map the geometric scalar into a high-dimensional continuous representation via a conditioning embedding function $\phi(R) \in \mathbb{R}^{d_r}$, where $d_r = 16$. The token embedding layer maps the discrete operator index $j_t$ to a vector $e_t \in \mathbb{R}^{d_e}$. The conditioning vector $\phi(R)$ is then concatenated with the token embedding to construct the augmented input representation $x_t$ for the subsequent Mamba layers:
\begin{equation}
    x_t = [e_t \,\|\, \phi(R)] \in \mathbb{R}^{d_e + d_r}
\end{equation}
where $\|$ denotes the concatenation operator. Following the formulation of Nakaji et al. \cite{nakaji2024gqe}, the model tracks the energy landscape by evaluating the log-probabilities of the sampled sequences. The predicted energy expectation value $\hat{E}(\vec{j} \mid R)$ of a generated circuit is computed as a linear projection of the cumulative hidden representations across the sequence length, matching the optimization alignment of the baseline while enforcing strict geometric awareness.

\subsection{Dynamic quality-threshold scheduling}
Candidate sequences are admitted to the replay buffer according to an energy-relative quality criterion. Rather than using a fixed threshold throughout training, Mamba-QE progressively tightens the acceptance criterion.

The scheduled threshold follows a linear annealing function,

\begin{equation}
\tau_k = \tau_{\mathrm{start}} +
\left(\tau_{\mathrm{end}}-\tau_{\mathrm{start}} \right) \frac{k}{K-1}
\end{equation}

where \(k\) is the current training iteration and \(K\) is the total number of training iterations. In the default configuration,

$$
\tau_{\mathrm{start}}=50\ {\rm mHa},
\qquad
\tau_{\mathrm{end}}=5\ {\rm mHa}.
$$

For a molecular geometry \(R\), a newly generated sequence with final energy \(E\) is accepted when

$$
E-E_{\mathrm{best}}(R)
\leq
\tau_k^{\mathrm{eff}}
$$

where \(E_{\mathrm{best}}(R)\) is the lowest energy previously observed for that geometry.

The implementation further introduces a self-healing mechanism to prevent premature starvation of the replay buffer. If the sequence acceptance rate falls below 5\% for two consecutive refinement events, the effective threshold is temporarily relaxed by 10 mHa. The accumulated relaxation is subsequently multiplied by 0.5 when the acceptance rate recovers. The effective threshold is bounded above by the initial 50 mHa value. This mechanism allows exploration to recover when an excessively restrictive threshold would otherwise prevent new sequences from entering the replay buffer.


\section{Results and Discusion}

The proposed Mamba Quantum Eigensolver (Mamba-QE) was evaluated on the hydrogen molecule (H$_2$), considering the prediction of the complete potential energy surface (PES) across 16 interatomic distances. Table I\ref{tab:h2_summary} summarizes the quantitative results obtained by the model.

The proposed architecture successfully reproduces the Full Configuration Interaction (FCI) reference energies with remarkable accuracy. The minimum energy predicted by Mamba-QE is $-1.136186$ Ha, differing from the exact FCI minimum of $-1.136189$ Ha by less than $3\times10^{-6}$ Ha, which corresponds to an error well below the chemical accuracy threshold of 1 kcal/mol (approximately 1.6 mHa). Moreover, chemical accuracy was achieved for all 16 molecular geometries evaluated, demonstrating that the proposed architecture generalizes consistently across the entire potential energy surface rather than only around the equilibrium geometry.

An important observation is that no variational violations were detected during the optimization process. Since the variational principle guarantees that approximate energies should not fall below the exact ground-state energy, the absence of violations indicates that the reinforcement learning procedure remained numerically stable while preserving the physical consistency of the generated quantum circuits.

The replay buffer statistics further illustrate the learning dynamics of the generative model. Although the mean energy stored in the final buffer ($-1.006522$ Ha) remains significantly above the ground-state energy, the best circuit found reaches $-1.134656$ Ha before the final refinement stage. This behavior suggests that the policy explores a diverse set of candidate circuits while progressively concentrating probability mass around lower-energy solutions. Such diversity is particularly desirable in reinforcement learning settings, preventing premature convergence to suboptimal circuit structures.

Unlike GPT-QE, whose Transformer backbone requires quadratic computational complexity with respect to the circuit length, the proposed Mamba-QE replaces the attention mechanism with a selective state-space model exhibiting linear complexity. Although the H$_2$ benchmark is too small to expose significant runtime differences, these results validate that replacing the Transformer does not degrade predictive accuracy. Therefore, the experimental evidence supports the hypothesis that Mamba-QE preserves the optimization capability of generative quantum eigensolvers while offering a substantially more scalable sequence modeling architecture for deeper quantum circuits and larger molecular systems.

Figures 1 and 2 further illustrate these findings. Figure 1 compares the predicted potential energy curve against the exact FCI reference, showing excellent agreement throughout the entire bond dissociation process. Figure 2 presents the Evolution of the Mamba-QE training loss for the H$_2$ molecule using the proposed dynamic scheduling strategy. 


\begin{figure}[H]
    \centering
    \includegraphics[width=0.9\linewidth]{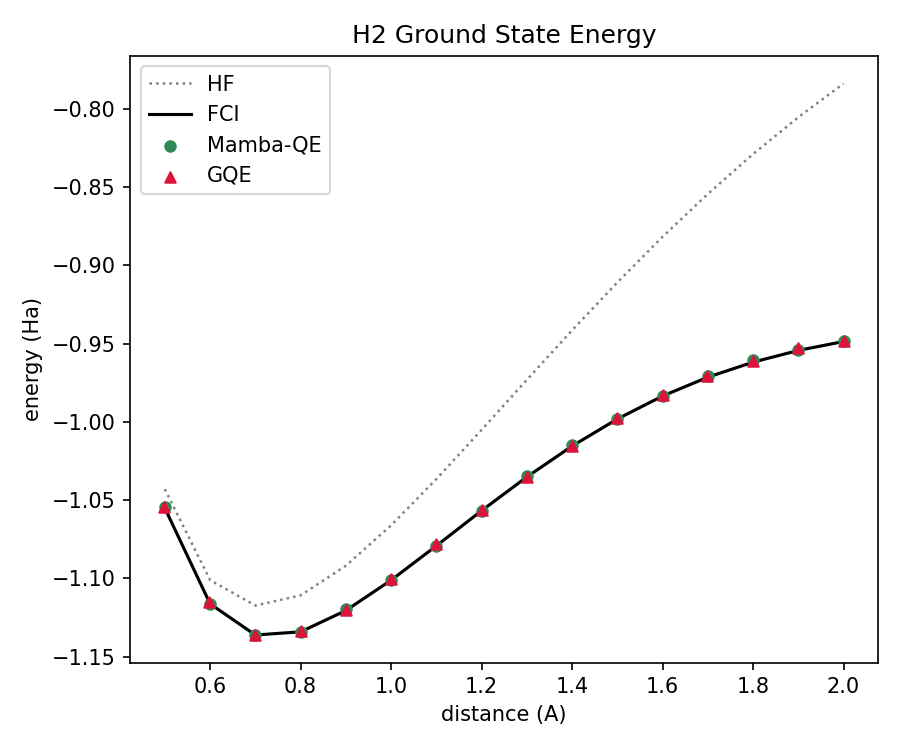}
    \caption{Potential energy surface of the $H_{2}$ molecule obtained with Mamba-QE and compared against the Full
    Configuration Interaction (FCI) and Hartree-Fock (HF) energy reference. The proposed architecture accurately reproduces the bond dissociation
    curve across all evaluated internuclear distances, achieving chemical accuracy throughout the sampled geometry
    range.}
    \label{fig:placeholder}
\end{figure}

\begin{figure}[H]
    \centering
    \includegraphics[width=0.9\linewidth]{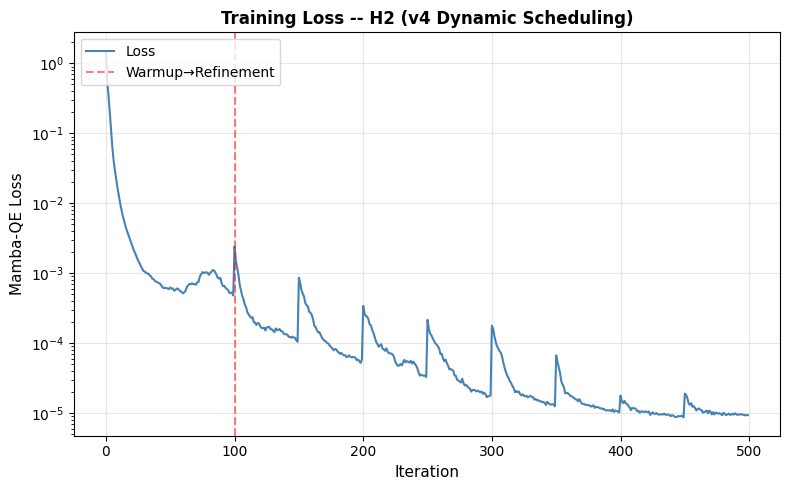}
    \caption{Training loss of the proposed Mamba-QE for the H$_2$ molecule under the dynamic scheduling strategy. The dashed vertical line indicates the transition from the supervised warm-up stage to the iterative refinement phase. After a rapid initial decrease, the loss exhibits periodic oscillations associated with successive policy update cycles while maintaining an overall downward trend, converging to values on the order of $10^{-5}$.}
    \label{fig:placeholder}
\end{figure}

\begin{figure}[H]
    \centering
    \includegraphics[width=0.9\linewidth]{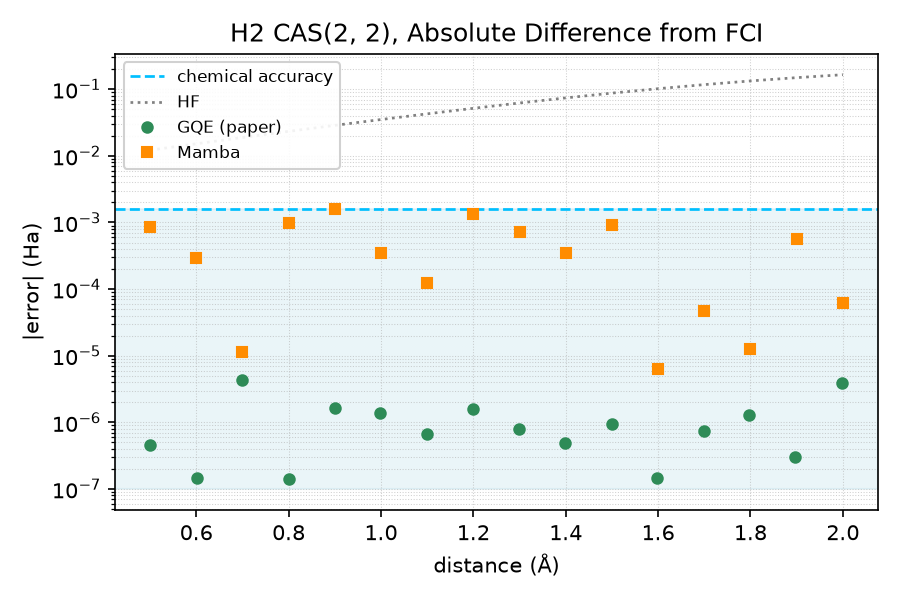}
    \caption{The absolute error from FCI. The chemical accuracy threshold (the blue dashed line) from $H_2$ in compare with GPT-QE (the green dot), Hartree-Fock method (black dashed line) and Mamba-QE (orange box).}
    \label{fig:placeholder}
\end{figure}

\begin{table}[H]
\centering
\caption{Summary of H$_2$ Results}
\begin{tabular}{lc}
\hline
\textbf{Quantity} & \textbf{Value} \\
\hline
FCI minimum (Ha) & $-1.136189$ \\
Bond length at FCI minimum (\AA) & $0.70$ \\
Mamba minimum (Ha) & $-1.136186$ \\
Error at minimum (mHa) & $0.00$ \\
Chemical accuracy & $16/16$ \\
Final buffer size & $129$ \\
Best buffer energy (Ha) & $-1.134656$ \\
Mean buffer energy (Ha) & $-1.006522$ \\
Distances covered & $10$ \\
Final threshold (mHa) & $5.0$ \\
Variational violations & $0/16$ \\
\hline
\end{tabular}
\label{tab:h2_summary}
\end{table}

\begin{table}[H]
\centering
\footnotesize
\setlength{\tabcolsep}{3pt}
\renewcommand{\arraystretch}{1.0}
\caption{Architectural comparison between GPT-QE and Mamba-QE}
\label{tab:h2_compare}
\begin{tabular}{l l l}
\hline\hline
\textbf{Component} & \textbf{GPT-QE} & \textbf{Mamba-QE} \\
\hline
Molecule & H$_2$ & H$_2$ \\
Basis set & sto-3g & sto-3g \\
Active space & CAS(2,2) & CAS(2,2) \\
Backbone & Transformer (GPT-2) & Selective SSM (Mamba) \\
Complexity & O(L$^2$) & O(L) \\
Sequence length & 10 & 4 \\
Hidden dimension & -- & 128 \\
State dimension & -- & 16 \\
Layers & 6 & 6 \\
Attention heads & 6 & -- \\
Expansion factor & -- & 2 \\
Convolution & -- & Kernel size 4 \\
Training & 200 epochs & 500 iterations \\
Iterations/epoch & 5 & -- \\
Samples/epoch & 50 & -- \\
Samples/distance & -- & 128 \\
Training geometries & 16 distances & 10 distances \\
Evaluation geometries & 16 distances & 16 distances \\
Conditioning & No & Yes (16-dim) \\
Loss & LM + GRPO & Cumulative MSE \\
Buffer & FIFO (1000) & Quality-threshold \\
Threshold schedule & -- & 50$\rightarrow$5 mHa \\
\hline\hline
\end{tabular}
\end{table}

\section{Conclusion}
In this work, we introduced the Mamba Quantum Eigensolver (Mamba-QE), a generative framework for ground-state molecular simulation that replaces the quadratic-complexity Transformer backbone of GPT-QE with a Selective State Space Model (SSM) exhibiting linear complexity with respect to sequence length. Furthermore, by incorporating a distance-conditioning mechanism, $\phi(R)$, Mamba-QE extends the Generative Quantum Eigensolver (GQE) paradigm from single-geometry optimization to the direct generation of quantum circuits across an entire potential energy surface (PES) within a single, unified policy.

Evaluated on the $H_2$ molecule, Mamba-QE reproduced Full Configuration Interaction (FCI) reference energies with chemical accuracy across all 16 sampled internuclear distances. The model achieved a minimum-energy error below $3\times 10^{-6}Ha$ and maintained zero variational violations throughout training. These results confirm that substituting the attention mechanism with a selective SSM preserves both the expressive power and the optimization stability of the generative approach, while successfully eliminating the $\mathcal{O}(L^2)$ scaling bottleneck that restricts Transformer-based GQE implementations as circuit depth and active-space sizes increase.

Although the $H_2$ benchmark rigorously validates the methodology and establishes a robust proof of concept, larger molecular systems are required to fully expose the runtime advantages inherent to the linear-complexity backbone. Future work will therefore extend Mamba-QE to larger, strongly correlated systems—such as transition-metal complexes and complex bond-breaking regimes—to directly quantify runtime scaling relative to GPT-QE. Additionally, we aim to investigate whether the distance-conditioning mechanism generalizes to multi-dimensional geometric parameter spaces beyond a single internuclear coordinate. Further research directions include exploring transfer learning across diverse molecular configurations, integrating the framework with hardware-executed circuit evaluations, and conducting deeper interpretability analyses of the operator sequences selected by the model to disentangle static and dynamic electronic correlation along the PES.

\section*{Funding}

This work was supported by EMBRAPII-QuIIN – Extension Courses in Quantum Technologies, through the Quantum Industrial Innovation (QuIIN) initiative at the EMBRAPII CIMATEC Competence Center in Quantum Technologies, with financial resources from the PPI IoT/Manufatura 4.0 of the MCTI, under grant number 053/2023, signed with EMBRAPII. C. Cruz acknowledge the  support from Fundação de Amparo à Pesquisa do Estado da Bahia - FAPESB under grant number BOL2809/2025 and CNPq by grant number 446761/2025-7.

\bibliographystyle{apsrev4-1}
\bibliography{ref}

\end{document}